\documentclass[aps,pra,twocolumn,superscriptaddress,floatfix,altaffilletter]{revtex4-1}

\usepackage[utf8]{inputenc}
\usepackage{amsmath,amsfonts,amssymb,amsthm}
\usepackage{graphicx}

\usepackage{subcaption}

\usepackage{xcolor}

\usepackage{siunitx}

\usepackage{tikz}
\usetikzlibrary{positioning}
\usepackage{pgfplots}
\pgfplotsset{compat=1.10}

\def\mathclap#1{\text{\hbox to 0pt{\hss$\mathsurround=0pt#1$\hss}}}
\newcommand{\im}{\mathrm{i}}
\renewcommand{\vec}[1]{\mathbf{#1}}

\begin{document}


\title{Neutral Helium Atom Diffraction from a Micron Scale Periodic Structure: Photonic Crystal Membrane Characterization}



\author{Torstein Nesse}
\thanks{T. Nesse and S. D. Eder contributed equally to this work}
\email{corresponding author torstein.nesse@ntnu.no}
\affiliation{Department of Physics, NTNU Norwegian University of Science and Technology, NO-7491 Trondheim, Norway}

\author{Sabrina D. Eder}
\thanks{T. Nesse and S. D. Eder contributed equally to this work}
\email{corresponding author torstein.nesse@ntnu.no}
\affiliation{Department of Physics and Technology, University of Bergen, All\'{e}gaten 55, 5007 Bergen, Norway}

\author{Thomas Kaltenbacher}
\affiliation{Department of Physics and Technology, University of Bergen, All\'{e}gaten 55, 5007 Bergen, Norway}

\author{Jon Olav Grepstad}
\affiliation{Tunable InfraRed Technologies AS, Gaustadalleen 21, 0349 Oslo, Norway}

\author{Ingve Simonsen}
\affiliation{Department of Physics, NTNU Norwegian University of Science and Technology, NO-7491 Trondheim, Norway}
\affiliation{Surface du Verre et Interfaces, UMR 125 CNRS/Saint-Gobain, F-93303 Aubervilliers, France}

\author{Bodil Holst}
\affiliation{Department of Physics and Technology, University of Bergen, All\'{e}gaten 55, 5007 Bergen, Norway}



\date{\today}

\begin{abstract}
	Surface scattering of neutral helium beams created by supersonic expansion is an established technique for measuring structural and dynamical properties of surfaces on the atomic scale. Helium beams have also been used in Fraunhofer and Fresnel diffraction experiments. Due to the short wavelength of the atom beams of typically \SI{0.1}{\nano\meter} or less, Fraunhofer diffraction experiments in transmission have so far been limited to grating structures with a period (pitch) of up to \SI{200}{\nano\meter}. However, larger periods are of interest for several applications, for example for the characterization of photonic crystal membrane structures, where the period is typically in the micron/high sub-micron range. Here we present helium atom diffraction measurements of a  photonic crystal membrane structure with a two dimensional square lattice of \num{100}~$\times$~\num{100} circular holes. The nominal period and hole radius were \SI{490}{\nano\meter} and \SI{100}{\nano\meter} respectively. To our knowledge this is the largest period that has ever been measured with helium diffraction. The helium diffraction measurements are interpreted using a model based on the helium beam characteristics. It is demonstrated how to successfully extract values from the experimental data for the average period of the grating, the hole diameter and the width of the virtual source used to model the helium beam.
\end{abstract}


\maketitle

\section{Introduction}
Helium atom scattering is a well-established technique in surface science. Elastic helium scattering is used to measure the structural properties of surfaces through diffraction and step height interference measurements. Inelastic helium  scattering is used to measure surface dynamics properties such as diffusion and vibrations. The advantage of helium scattering lies in the very low energy of the beam (typically less than \SI{0.1}{\electronvolt}) and the fact that the beam is neutral, which means that it is possible to investigate insulating and/or fragile surfaces and adsorbates. The small wavelength of the helium beam (less than \SI{0.1}{\nano\meter}) means that it is very well suited for investigating structures on the atomic scale. Larger scale structures put severe demands on the beam collimation and angular resolution of the diffraction system. The largest period of a surface periodic structure that has been resolved using helium scattering is, to our knowledge, a surface reconstruction of $\alpha$-quartz~(0001) with a period of \SI{5.55}{\nano\meter}~\cite{Eder2015-2}. For reviews of the use of helium atom surface scattering in surface science see Refs.~\cite{Farias1998,Holst.2013}.

Transmission helium atom diffraction has so far been used only in a limited number of experiments. This is mainly due to the fact that the low energy of the helium atoms means that they do not penetrate any solid materials and hence transmission experiments can only be performed on porous structures. Experiments have been carried out on grating structures with a period of up to \SI{200}{\nano\meter}~\cite{Schoellkopf1994,Grisenti1999}, and various Fresnel diffraction and focusing experiments have been done using zone plates and a Poisson spot aperture~\cite{PhysRevLett.83.4229,koch08,ThomasReisinger2009,Eder2015}. In this paper we present the first experiment carried out on a two-dimensional lattice structure: A photonic crystal membrane structure with a nominal period of \SI{490}{\nano\meter} and a hole radius of \SI{100}{\nano\meter}. 

Photonic crystals have a number of potential applications~\cite{Joannopoulos1997}. They have been demonstrated as building blocks in integrated circuits~\cite{Kuramochi2014}, can be used as mirrors in applications requiring very high optical power or operating temperatures above a couple of hundred degrees Celsius~\cite{Jeong2013}, and are applied in the fabrication of quantum dots and quantum light sources applicable in quantum computing~\cite{Lodahl2015}. The commercial use of photonic crystals is currently limited to transducer elements in biosensors~\cite{Cunningham2016} and to maximize the light extraction efficiency of light emitting diodes (LED)~\cite{Matioli2010}. The particular type of photonic crystal structure investigated in this paper was developed for the detection of single molecules~\cite{grepstad12}, and so it was particularly important to know if it was really transparent to atoms. This can be difficult to determine from a scanning electron microscope image, since the electrons may still penetrate nanometer thick residue layers making the sample appear transparent in a region where it in reality is not.

\section{\label{sec:experimental_setup}Experimental setup}

\begin{figure}
	\centering
	\includegraphics[width=\columnwidth]{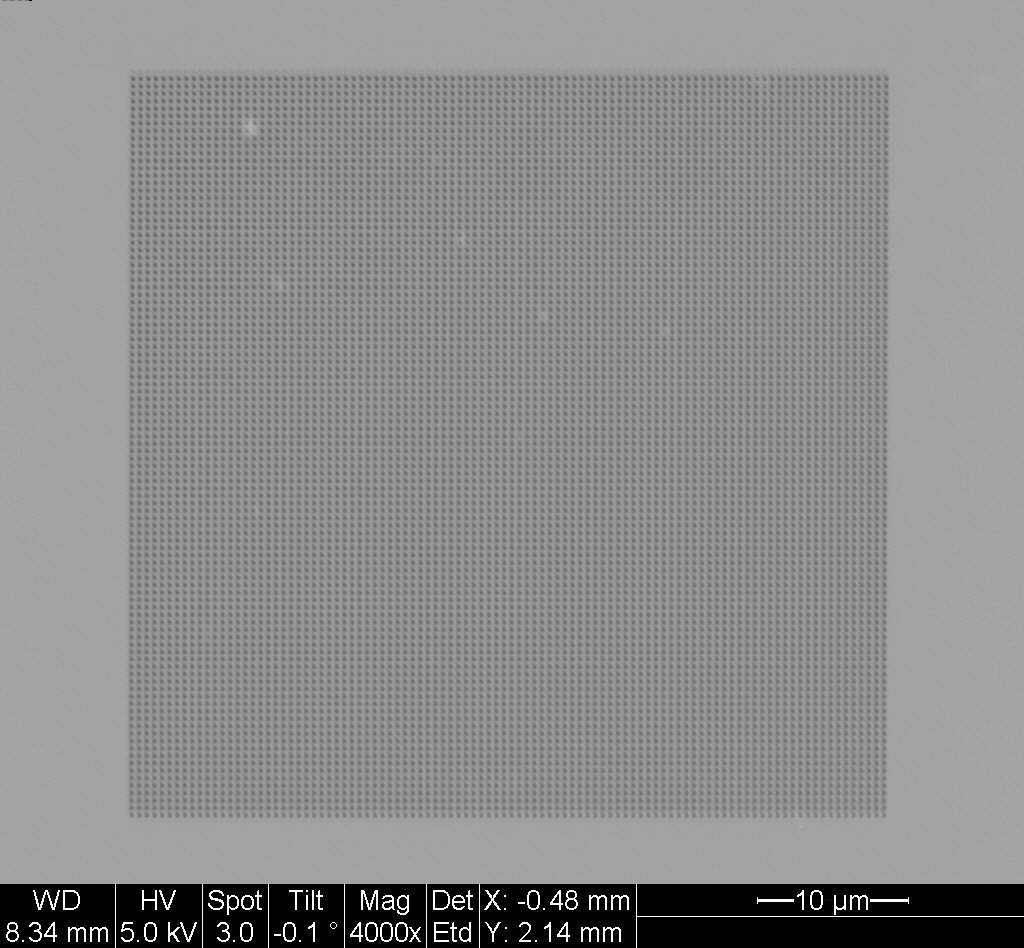}
	\caption{\label{fig:crystal}Scanning electron microscopy image of the photonic crystal sample. The structure consists of \num{100}~$\times$~\num{100} holes on a square grid with a period of \SI{490}{\nano\meter}. The holes were circular each with a diameter of \SI{200}{\nano\meter}.}
\end{figure}

The photonic crystal membrane sample characterized in these experiments was composed of an array of through holes in a \SI{700}{\micro\meter}~$\times$~\SI{700}{\micro\meter} free-standing membrane, suspended in a silicon frame. The thickness of the membrane was \SI{150}{\nano\meter} and made of silicon nitride and silicon oxide thin films (\SI{50}{\nano\meter} Si3N4 - \SI{50}{\nano\meter} SiO2 - \SI{50}{\nano\meter} Si3N4). The holes were made using electron beam lithography combined with reactive ion etching. The pattern consisted of a square array of \num{100}~$\times$~\num{100} circular holes. Each hole had a radius of \SI{100}{\nano\meter} and the grid had a period of \SI{490}{\nano\meter}, giving a \SI{50}{\micro\meter}~$\times$~\SI{50}{\micro\meter} patterned area containing a total of \num{10000} holes. A SEM image of the crystal is presented in Fig.~\ref{fig:crystal}. For a detailed description of the preparation see Ref.~\cite{grepstad12}.

\begin{figure*}
	\centering
\begin{tikzpicture}

\def\sourcex{-5};
\def\skimmerx{-4};
\def\aponex{-1.6};
\def\aptwox{0.8};
\def\apthreex{3};
\def\detx{4};

\def\gridtop{2}
\def\labelline{2.3}
\def\labelpoint{1.8}

\def\dstoff{-0.2}

\draw [black!20, fill=black!20] ({\skimmerx-0.4},-0.1) -- (0,-0.5) -- (0,0.5) -- ({\skimmerx-0.4},0.1);
\draw [black!20, fill=black!20] (0,-0.2) -- (\apthreex,-0.8) -- (\apthreex,0.8) -- (0,0.2);
\draw [black!20, fill=black!20] (\detx,-0.2) -- (\apthreex,-0.1) -- (\apthreex,0.1) -- (\detx,0.2);

\draw [thick, black, fill=black!20] ({\sourcex-0.5},-0.2) -- (\sourcex,-0.2) -- (\sourcex,0.2) -- ({\sourcex-0.5},0.2) -- ({\sourcex-0.5},-0.2);

\draw [black!20, fill=black!20] ({\sourcex},0.1) -- ({\sourcex-0.1},0.1) -- ({\sourcex-0.1},-0.1) -- ({\sourcex},-0.1);
\draw [black!20, fill=black!20]
({\sourcex},0.1) to [out=30, in=70, looseness=2]
({\skimmerx-0.4},0.1) -- 
({\skimmerx-0.4},0.0) -- 
({\skimmerx-0.4},-0.1) to [out=290, in=330, looseness=2] 
({\sourcex},-0.1);

\draw[thick] (\skimmerx,{-\gridtop}) -- (\skimmerx,-0.4) -- ({\skimmerx-0.4},-0.1);
\draw[thick] ({\skimmerx-0.4}, 0.1) -- (\skimmerx, 0.4) -- (\skimmerx,\gridtop);

\draw[thick] (\aponex,{-\gridtop}) -- (\aponex,-0.8);
\draw[thick] (\aponex,0.8) -- (\aponex,\gridtop);

\draw[thick] (0,{-\gridtop}) -- (0,-0.2);
\draw[thick, dotted] (0,-0.2) -- (0,0.2);
\draw[thick] (0,0.2) -- (0,\gridtop);

\draw[thick] (\aptwox,{-\gridtop}) -- (\aptwox,-0.4);
\draw[thick] (\aptwox,0.4) -- (\aptwox,\gridtop);

\draw[thick] (\apthreex,{-\gridtop}) -- (\apthreex,-0.1);
\draw[thick] (\apthreex,0.1) -- (\apthreex,\gridtop);


\draw[dashed, black!50] ({\sourcex-0.5},0) -- (\apthreex,0);

\draw [] ({\sourcex-0.25},0.2) -- ({\sourcex-0.3},1) node [above] {nozzle};
\draw [] (0,\labelpoint) -- (-0.3,{\labelline-0.058}) node [above] {sample};
\draw [] (\aponex,\labelpoint) -- ({\aponex-0.5},\labelline) node [above] {AP 1};
\draw [] (\skimmerx,\labelpoint) -- ({\skimmerx-0.5},\labelline) node [above] {skimmer};
\draw [] (\aptwox,\labelpoint) -- ({\aptwox+0.3},\labelline) node [above] {AP 2};
\draw [] (\apthreex,\labelpoint) -- ({\apthreex-0.2},{\labelline-0.058}) node [above, text width=1cm,align=center] {scanning\\plane};

\draw[thick, <->] ([shift=(50:0.8)]{\apthreex-0.1},0) arc (50:-50:0.8) node[below]{$\theta$};


\draw[thick, |-]
({\skimmerx-0.4},{-\gridtop+\dstoff}) node[below]{$x_3^*$}
-- node[below]{$z_{ss}$}
(0,{-\gridtop+\dstoff}) node[below]{$x'_3$};
\draw[thick, |-|]
(0,{-\gridtop+\dstoff})
-- node[below]{$z$}
(\apthreex,{-\gridtop+\dstoff}) node[below]{$x_3$};

\draw[thick, <->]
({\sourcex-0.5},{-\gridtop+\dstoff+0.5}) node[left]{$\hat{\vec{x}}_1$}
-- ({\sourcex-0.5},{-\gridtop+\dstoff}) --
({\sourcex},{-\gridtop+\dstoff}) node[below]{$\hat{\vec{x}}_3$};

\begin{axis}[axis lines=none, anchor=origin, rotate around={-90:(current axis.origin)}, width=40ex, height=25ex, at={(28ex, 0)}]
\addplot[black, thick] coordinates {
(-0.0004526155651453721, 475.13229090000004)
(-0.00044691464143120294, 471.27236020000004)
(-0.00044121371774368684, 465.8148523)
(-0.00043551279410070467, 472.3976358)
(-0.000429811870520138, 471.69853050000006)
(-0.0004241109469287617, 472.3086721000001)
(-0.0004184100233626783, 484.40437199999997)
(-0.00041270909963933545, 472.6876305)
(-0.000407008176122818, 475.7880926)
(-0.0004013072526305734, 475.5105641)
(-0.00039560632916226174, 486.2429425999999)
(-0.00038990540571754307, 486.91672170000015)
(-0.00038420448229607745, 485.3284049)
(-0.0003785035588975248, 475.38826950000004)
(-0.00037280263552154515, 488.19559449999997)
(-0.00036710171216779847, 481.33100780000007)
(-0.0003614007888359448, 483.8389469)
(-0.00035569986552564417, 490.6835464)
(-0.0003499989422365567, 483.97720539999995)
(-0.0003442980187861298, 490.2578129)
(-0.0003385970955384484, 487.15866850000003)
(-0.00033289617231095994, 485.55111789999995)
(-0.0003271952491033247, 492.6828302)
(-0.0003214943259152026, 483.19850180000003)
(-0.00031579340274625353, 491.3357682000001)
(-0.00031009247959613764, 481.0667568000001)
(-0.0003043915564645148, 485.03494610000007)
(-0.0002986906333510452, 479.8952537)
(-0.0002929897102553887, 485.14982370000007)
(-0.00028728878717720544, 489.22426390000004)
(-0.00028158786411615534, 487.9819717)
(-0.000275886940889686, 485.6296333)
(-0.00027018601731524516, 483.62900940000003)
(-0.00026448509466797936, 489.76947240000004)
(-0.00025878417203648686, 496.8389273000001)
(-0.0002530832494204276, 500.06417280000005)
(-0.0002473823268194616, 492.68373460000004)
(-0.00024168140241112498, 501.52176090000006)
(-0.00023598047983932545, 516.3241786)
(-0.0002302795572815992, 507.86955800000015)
(-0.00022457863473760632, 515.8019398)
(-0.0002188777103848829, 524.5252389999999)
(-0.00021317678786733656, 535.8564977999999)
(-0.00020747586536250362, 544.2758854)
(-0.00020177494287004395, 547.3147486)
(-0.00019607402038961764, 537.8573012)
(-0.0001903730960987609, 544.1139054)
(-0.00018467217364138137, 543.4712398)
(-0.00017897125119501516, 551.8896589999999)
(-0.0001732703287593224, 551.5303312)
(-0.000167569406333963, 543.477767)
(-0.00016186848209647327, 549.5202084000001)
(-0.00015616755969076075, 548.4498841000001)
(-0.00015046663729436163, 543.7477481999999)
(-0.000144765714906936, 555.0740732999999)
(-0.0001390647925281438, 548.7142557000001)
(-0.00013336386833552133, 527.7188776000002)
(-0.00012766294597297607, 526.6122655)
(-0.00012196202361804435, 529.5167409)
(-0.00011626110127038614, 524.6255552)
(-0.00011056017892966149, 532.9775786)
(-0.00010485925477340652, 520.7141151)
(-9.915833244552887e-5, 509.42517469999996)
(-9.345741012356484e-5, 519.6195090999998)
(-8.775648780717432e-5, 509.7094383000001)
(-8.20555636738936e-5, 515.7520073)
(-7.635464136763025e-5, 519.5319404)
(-7.065371906592055e-5, 524.8782205000001)
(-6.495279676842439e-5, 529.4944235)
(-5.925187447480186e-5, 549.5764349)
(-5.355095036258926e-5, 542.6934378000001)
(-4.7850028075694e-5, 557.3353433999999)
(-4.214910579165241e-5, 572.1884736999999)
(-3.644818351012447e-5, 569.9488774999999)
(-3.0747261230770206e-5, 569.3399238000001)
(-2.504633713112594e-5, 590.7394162)
(-1.934541485509907e-5, 601.9102958)
(-1.3644492580225915e-5, 592.1983792000001)
(-7.943570306166485e-6, 585.251713)
(-2.242648032580849e-6, 603.0678831)
(3.4582760629948666e-6, 593.7216401000002)
(9.15919833665306e-6, 597.7178466)
(1.4860120610857424e-5, 588.8669081)
(2.0561042885948074e-5, 590.5942899)
(2.6261965162264927e-5, 576.9914720999999)
(3.196288926227174e-5, 575.3745268)
(3.766381154206092e-5, 561.1548446)
(4.336473382409633e-5, 566.3801898)
(4.9065656108717905e-5, 551.7116961)
(5.476658021838934e-5, 550.1258449000001)
(6.0467502509203216e-5, 544.0623189000002)
(6.616842480362322e-5, 541.8776906)
(7.186934710198931e-5, 520.6871324)
(7.757026940464156e-5, 518.4667723)
(8.327119353404358e-5, 521.8160534)
(8.8972115846288e-5, 516.4406272)
(9.467303816383846e-5, 516.1620951000001)
(0.00010037396048703501, 518.2998058000001)
(0.00010607488281621752, 519.7914131)
(0.00011177580697384995, 515.4897684000001)
(0.00011747672931602458, 525.7812031000001)
(0.00012317765166520517, 530.3141253)
(0.00012887857402173192, 523.3120402)
(0.00013457949638594451, 534.9902606)
(0.00014028042058030691, 538.1860159)
(0.00014598134296091157, 544.5913429999999)
(0.0001516822653502221, 541.3563021000001)
(0.0001573831877485787, 545.9661991000002)
(0.0001630841101563211, 546.6801522)
(0.00016878503439591322, 543.3250455000001)
(0.00017448595682344756, 558.9000438)
(0.00018018687926138775, 543.6848145)
(0.00018588780171007377, 549.5415326)
(0.0001915887259919696, 549.8954006)
(0.00019728964846316743, 546.3915380000001)
(0.00020299057094613104, 551.1165866)
(0.00020869149344120066, 531.2449513)
(0.00021439241594871597, 538.7506841000001)
(0.00022009334029114095, 518.2322299000001)
(0.00022579426282456807, 526.9536890000002)
(0.0002314951853714608, 525.3417318999999)
(0.00023719610793215948, 512.1356247)
(0.0002428970305070038, 504.1117109)
(0.0002485979549184577, 504.21411190000003)
(0.00025429887752261373, 502.7669239)
(0.00025999980014193525, 487.2836707000001)
(0.0002657007227767626, 490.667732)
(0.00027140164542743583, 488.618684)
(0.0002771025699164184, 478.1234880000001)
(0.00028280349259980305, 477.7782724000001)
(0.00028850441530005316, 484.6270104)
(0.00029420533801750897, 483.99663870000006)
(0.00029990626075251063, 487.8029185)
(0.00030560718532752167, 483.03217370000004)
(0.0003113081080986344, 484.1568088)
(0.000317009030888313, 483.00385689999996)
(0.000322709953696897, 479.41151840000003)
(0.0003284108783468506, 490.4261081000001)
(0.00033411180119426593, 486.48846480000003)
(0.0003398127240616068, 484.5397244)
(0.00034551364694921337, 481.749139)
(0.0003512145698574254, 486.3127454)
(0.0003569154946087068, 492.4970125000001)
(0.00036261641755915003, 483.17216360000003)
(0.0003683173405312186, 475.11309819999997)
(0.0003740182635252528, 482.43135570000004)
(0.0003797191865415925, 482.7878591)
(0.0003854201114027016, 479.4339782)
};
\end{axis}

\end{tikzpicture}
	\caption{\label{fig:tikz_meas_setup_scheme}Top view diagram of the experimental setup. The beam is created by supersonic expansion through a nozzle and the central part of the beam is selected by a conically shaped aperture (skimmer). Apertures AP1 and AP2 further collimate the beam. The sample is inserted in the beamline and the diffraction pattern is obtained by scanning the detector with a vertical slit horizontally across the pattern while the sample is kept still. See main text for further details.}
\end{figure*}
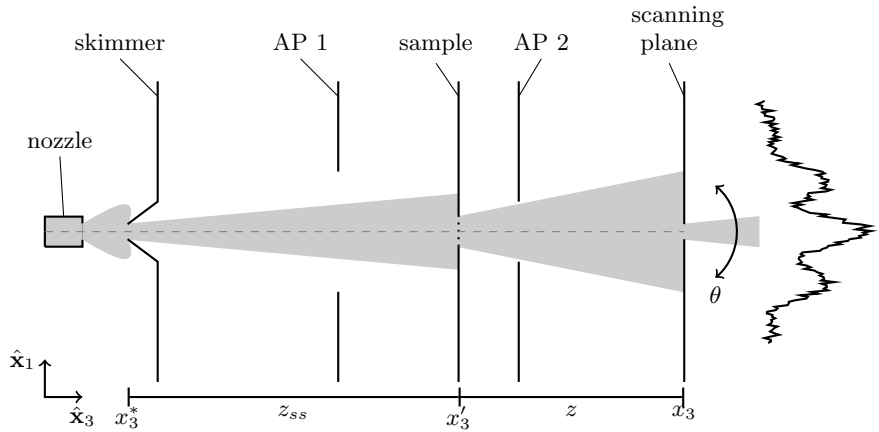

The diffraction  measurements were carried out using  the molecular beam apparatus MAGIE~\cite{Apf2005}. A diagram of the experimental setup is presented in Fig.~\ref{fig:tikz_meas_setup_scheme}. The helium beam was created by supersonic expansion through a \SI{10}{\micro\meter} diameter nozzle and collimated with a \SI{400}{\micro\meter} wide circular skimmer. Measurements were carried out using both a room temperature and a cooled beam, both with a stagnation pressure of \SI{91}{\bar}. The former had a beam temperature of \SI[separate-uncertainty = true]{296+-3}{\kelvin}, corresponding to a wavelength of \SI{0.57}{\angstrom} with a spread of \SI{+-0.015}{\angstrom}. The cooled beam had a beam temperature of \SI[separate-uncertainty = true]{126+-3}{\kelvin}, corresponding to a wavelength of \SI{0.87}{\angstrom} with a spread of \SI{+-0.015}{\angstrom}. The photonic crystal sample was placed in the beamline a distance $z_{ss}=$\SI[separate-uncertainty = true]{1528+-5}{\milli\meter} after the skimmer. A \SI{10}{\micro\meter} wide vertical slit was placed a distance $z=$\SI[separate-uncertainty = true]{1044+-5}{\milli\meter} after the sample just in front of the detector. This slit was then moved horizontally in steps of approximately \SI{5.7}{\micro\meter} along the $\hat{\vec{x}}_1$ direction to scan the diffraction pattern.

Care was taken to align the sample so that it was perpendicular to the incident beam and to align one of the axes of the photonic crystal with the scan direction $\hat{\vec{x}}_1$. We estimate that the alignment is precise to within \SI{+-0.5}{\degree}. To test the overall configuration the sample was rotated \SI{45}{\degree} and \SI{90}{\degree} around its normal from the initial alignment and additional room temperature measurements were performed. To reduce background contributions to the measured signal, two collimating apertures were inserted before and after the sample at distances \SI[separate-uncertainty = true]{566+-5}{\milli\meter} (AP1, \SI{400}{\micro\meter} diameter) and \SI[separate-uncertainty = true]{282+-5}{\milli\meter} (AP2, \SI{200}{\micro\meter} diameter) from the sample. The second aperture AP2 was moved along with the slit in front of the detector when scanning. Even with these precautions the signal to background ratio was still not optimal. For the different measurements performed at \SI{126}{\kelvin}, the maximum count rate was \SI[separate-uncertainty = true]{612.5+-9.7}{\per\second} and the minimum count rate was \SI[separate-uncertainty = true]{440.6+-2.6}{\per\second}, which corresponds to the strength of the fundamental peak and the background respectively. The final diffraction patterns presented here were made by averaging over a total of \num{10} individual scans across the diffraction pattern.

\section{\label{sec:diffraction_model}The diffraction model}

\subsection{\label{sec:diffraction_grating}Diffraction grating}
Since the distances between the source and the sample, and between the sample and the detector are very large compared to the size of the grid and the wavelength of the beam, the Fraunhofer approximation is valid and we can model the propagation of the helium beam as a scalar wave. We look at the propagation of the scalar beam from a plane sample to a scanning plane parallel to the sample. In the Fraunhofer approximation, the diffracted field at position $\vec{x}$ behind a single circular hole centered at position $\vec{x'}$ can be described by~\cite{Goodman05}
\begin{subequations}\label{eq:fraunhofer_circular_hole}
\begin{equation}
	\psi_\text{circ}(\vec{x}|\vec{x'}) = 
	\frac{4\pi A}{\im k z}
	e^{\im kz}e^{\im \frac{kr^2}{2z}}
	\left[
	\frac{J_1(kwr/z)}{kwr/z}
	\right] \psi_\text{inc}(\vec{x'}),
\end{equation}
where the  in-plane distance between the center of the hole and the observation position is defined
\begin{equation}
r=|\vec{x}_\|-\vec{x}_\|'|.
\end{equation}
\end{subequations}
In this equation $k=2\pi/\lambda$ is the incident wavenumber, $\lambda$ is the wavelength, $w$ is the radius of the hole, $z = (x_3-x_3')$ is the distance from the sample plane to the scanning plane, $A$ is the area of the hole and $\psi_\text{inc}(\vec{x'})$ is the incident field at the sample. A position in-plane with a normal in the propagation direction $\hat{\vec{x}}_3$ is denoted $\vec{x}_\| = (x_1,x_2,0)$. The incident field $\psi_\text{inc}(\vec{x'})$ changes across the sample, but we assume that it is approximately constant across a single hole.

To model the result of the diffraction grating we take the superposition of the fields found using Eq.~\eqref{eq:fraunhofer_circular_hole} for each hole in the two-dimensional grid of $N~\times~N$ holes:

\begin{equation}
	\psi(\vec{x}) = \sum_{\vec{n}} \psi_\text{circ}(\vec{x}|\vec{x'}_{\vec{n}}),
\end{equation}
where $\vec{n}=(n_1,n_2)$ is the hole coordinate in the grid ($n_i \in [1,N]$), and $\vec{x'}_{\vec{n}}$ is the hole position.

To relate the field $\psi(\vec{x})$ to the measured quantity we glide a slit over the intensity in the scanning plane $I(\vec{x})~=~|\psi(\vec{x})|^2$, and integrate over the intensity inside the slit at each position. This leads to a quantity similar to the one captured in the experiment

\begin{equation}
	I_\text{sim}(x_d)
	=
	\int\limits^{x_d+\frac{w_s}{2}}_{x_d-\frac{w_s}{2}}\mathrm{d}x_1
	\int\limits^{h_s}_{-h_s}\mathrm{d}x_2
	\left\{\left| \psi(\vec{x})\right|^2\right\}\biggr\rvert_{x_3 = z},
	\label{eq:I_sim}
\end{equation}
where $x_d$ is the center position of the slit along the scanning direction $\hat{\vec{x}}_1$,  and $w_s$ and $h_s$ is the width and the height of the scanning slit respectively.

\subsection{\label{sec:source_model}Source description}
The supersonic source gives rise to an incoherent beam of helium atoms with a narrow speed distribution. In principle, the speed distribution will cause the diffraction pattern to ``smear out'' due to the difference in wavelength; however, in the experimental results that we will present, the speed distribution is so narrow that this effect is not very prominent. A change in wavelength equal to the spread given in Sec.~\ref{sec:experimental_setup} shifts the position of the first order diffraction peaks by approximately \SI{3}{\micro\meter}. Hence, in the following, the beam will be assumed to be characterized by the wavelength that corresponds to its center energy (or speed).

To describe the Helium source used in the experiments we will adapt the \textit{virtual source model} that was introduced by Beijerinck and Verster to describe supersonic expansions~\cite{Beijerinck1981}. Here, the atoms initially collide until they eventually reach the molecular flow regime at a distance from the nozzle referred to as the quitting surface. When this happens, the individual trajectories can be traced back to a plane that is perpendicular to the mean direction of travel and where the width of the spatial distribution function of the trajectories is at a minimum --- the virtual source. This spatial distribution can be fitted with one or two Gaussian functions~\cite{Beijerinck1981,DePonte2006}. 

Within the virtual source model, the incident beam is considered as an \textit{incoherent} and weighted \textit{superposition} of spherical waves (point sources) located approximately in the skimmer plane. Here, the weight (or amplitude) used in the superposition will be taken to be a Gaussian function whose width, called $\sigma$ below, mimics the half width of the skimmer. Mathematically the incident field at position $\vec{x}'$ can therefore be written in the form
\begin{equation}
	\psi_\text{inc}(\vec{x'}) 
	=
	\int\limits\mathrm{d}^2\!x^*_\| \,
	\frac{e^{-\frac{{x^*_\|}^2}{2\sigma^2}}}{\sqrt{2\pi\sigma^2}}
	\frac{e^{\im k |\vec{x'}-\vec{x^*}|}}{|\vec{x'}-\vec{x^*}|}
	e^{\im \phi(\vec{x^*_\|})},
	\label{eq:source_description}
\end{equation}
where $\vec{x}^*_\|$ denotes a position in the skimmer plane and $\phi(\vec{x}^*_\|)$ represents a random phase function associated with the spherical wave at $\vec{x}^*_\|$. This function is assumed to be an uncorrelated stochastic variable that is uniformly distributed on the interval $[0,2\pi\rangle$. The incident amplitude has been set to $1$ in Eq.~\eqref{eq:source_description}.

To perform simulations using the incident field, the integral in Eq.~\eqref{eq:source_description} has to be evaluated numerically and the results that depend on it are averaged over an  ensemble of realizations of the random phase function.

It will be seen in Sec.~\ref{sec:results} that the form of the incident field given by Eq.~\eqref{eq:source_description} is sufficient to explain the measured results.

\subsection{Fit to experimental data}
A rescaling is necessary to compare the results of the diffraction model described in Secs.~\ref{sec:diffraction_grating} and~\ref{sec:source_model} with the experimental data. The experimental data are captured as counts per second, which we must scale the simulation data to fit. The experimental data also contains a strong background signal. To take these effects into account we have fitted the results from the diffraction model to the experimental data using two variables; one for scaling the overall intensity to match the source intensity $\alpha$ and one for shifting the results and taking the background into account $I_\text{0}$,

\begin{equation}
	\langle I(x_d) \rangle = \alpha \langle I_\text{sim}(x_d) \rangle +I_\text{0}.
	\label{eq:experimental_fit}
\end{equation}
Here $\langle\cdot\rangle$ is the mean over several realizations of the random phase function $\phi(\vec{x}^*)$ in Eq.~\eqref{eq:source_description}. The best values for $\alpha$ and $I_\text{0}$ were then found by a least squares fit of $I(x_d)$ to the experimental data $I_\text{exp}(x_d)$. The fit was performed using the error norm
\begin{equation}
	\left\lVert I_\text{exp}(x_d) - \langle I(x_d) \rangle \right\rVert_2 = \sqrt{\sum_{x_d} \frac{|I_\text{exp}(x_d) - \langle I(x_d) \rangle|^2}{\sigma_\text{exp}^2}},
	\label{eq:error_norm}
\end{equation}
where $\sigma_\text{exp}$ is the standard error of the experimental measurements.

\section{\label{sec:results}Results and Discussion}

%
\begin{figure}
\centering
\includegraphics[width=\columnwidth]{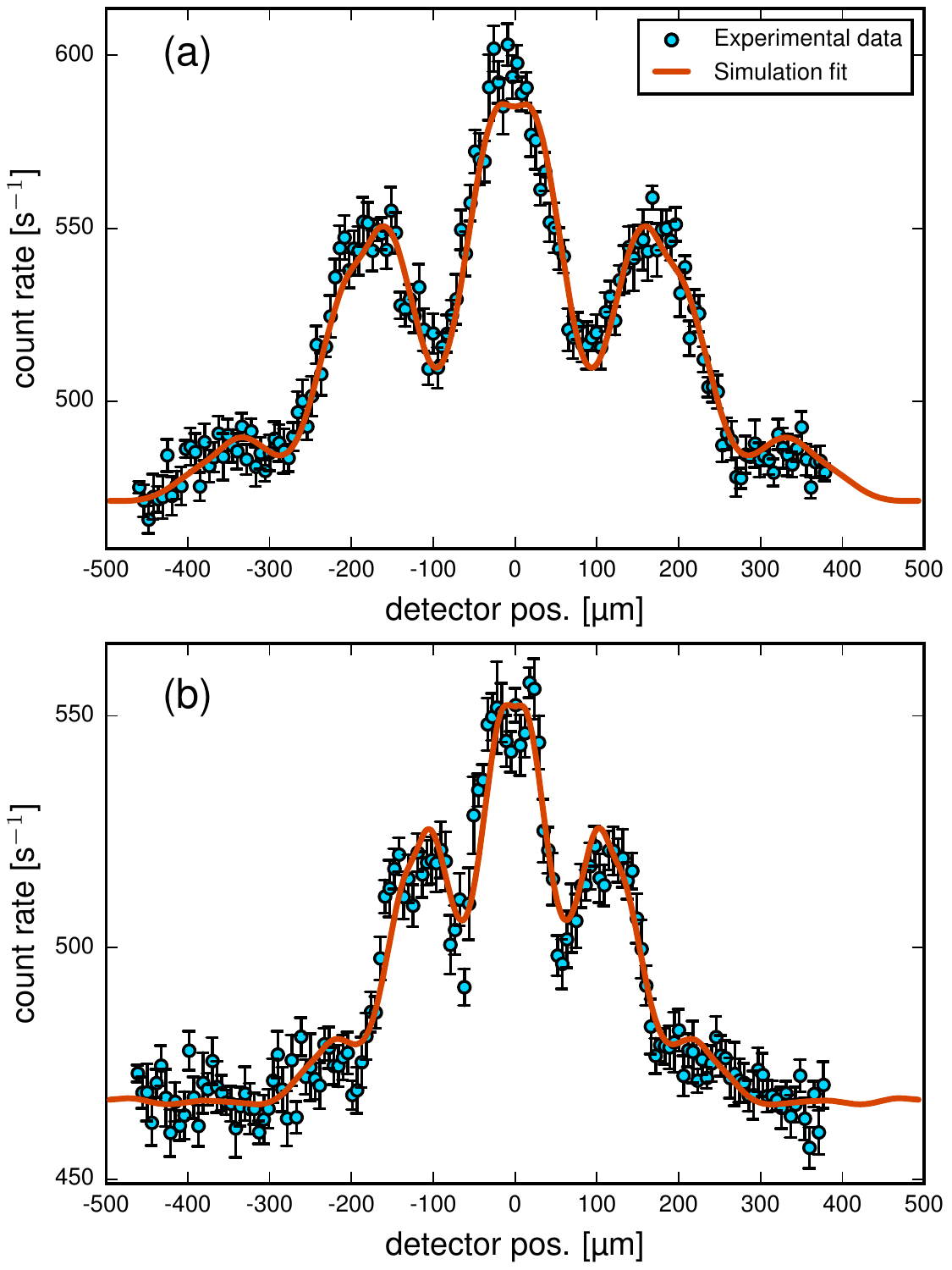}
\caption{\label{fig:120K_300K}(Color online) Measured helium diffraction intensities (symbols) transmitted through the sample depicted in Fig.~\ref{fig:crystal} as a function of the horizontal detector position $x_d$ when using a vertical slit of width $w_s=$\SI{10}{\micro\meter} in front of the detector. The temperature of the incident helium beam was (a) \SI{126}{\kelvin} and (b) \SI{296}{\kelvin}. The sample was aligned so that one of the lattice vectors of the square grating was parallel with $\mathbf{\hat{x}}_1$. The solid lines represents the best fits obtained on the basis of Eqs.~\eqref{eq:I_sim}, \eqref{eq:experimental_fit} and \eqref{eq:error_norm} when assuming the experimental parameters presented in Table~\ref{tab:simulation_parameters}. The source widths $\sigma$ for the two temperatures were (a)~\SI{77}{\micro\meter} and (b)~\SI{52}{\micro\meter}. The simulation results presented are the  mean of the results for \num{1000} source realizations.}
\end{figure}
%

\begin{table}
	\caption{\label{tab:simulation_parameters}Parameters used for the simulation results shown in Figs.~\ref{fig:120K_300K}, \ref{fig:300K_rotation}, \ref{fig:300K_45} and \ref{fig:rsd_scan}. All of the parameters are based on measurements on the experimental setup.}
	\begin{ruledtabular}
	\begin{tabular}{lcc}
		Parameter    & \SI{126}{\kelvin} & \SI{296}{\kelvin} \\
		\hline
		Wavlength $\lambda$ [\si{\angstrom}]& \num{0.87} & \num{0.57} \\
		Source to sample distance $z$ [\si{\meter}]& \multicolumn{2}{c}{\num{1.5284}} \\
		Skimmer radius [\si{\micro\meter}]& \multicolumn{2}{c}{\num{200}} \\
		Hole radius $w$ [\si{\nano\meter}] & \multicolumn{2}{c}{\num{100}} \\
		Hole periodicity $p$ [\si{\nano\meter}] & \multicolumn{2}{c}{\num{490}} \\
		Sample to scan slit distance $z_{ss}$ [\si{\meter}]& \multicolumn{2}{c}{\num{1.044}} \\
		Slit size $w_s~\times~h_s$ & \multicolumn{2}{c}{\SI{10}{\micro\meter}~$\times$~\SI{1}{\milli\meter}} 
	\end{tabular}
	\end{ruledtabular}
\end{table}

Helium diffraction intensity measurements in transmission were performed for beam temperatures \SI{126}{\kelvin} and \SI{296}{\kelvin}, and the results are presented as circles in Figs.~\ref{fig:120K_300K}(a) and \ref{fig:120K_300K}(b), respectively. 
The presented measurements are the arithmetic mean of \num{10} independent horizontal detector scans ($x_1$-scans) for each beam temperature. The error bars reported for each experimental data point correspond to the standard deviation on the mean calculated for each point. 

The measurements presented in Fig.~\ref{fig:120K_300K} show pronounced diffraction patterns. The count rates for the zeroth and first order diffraction peaks are all more than \SI{10}{\percent} higher than the background level. The observed diffraction patterns behave as expected: They are symmetric around the location of the fundamental order peak located at $x_1=0$, and the positions and widths of the diffraction peaks vary with temperature. In particular, one observes that when the temperature of the incident beam is increased, the distance between the positions of the fundamental and the first diffractive orders and the widths of the peaks both become smaller. Since the wavelength associated with the beam of incidence is inversely proportional to the temperature of the beam (see Table~\ref{tab:simulation_parameters}), such behavior is expected from the grating equation from physical optics~\footnote{Under the assumption of normal incidence, the grating equation (in 1D for simplicity) reads $\sin\theta_t=n\lambda/a$ where $\theta_t$ denotes the transmission angle of the diffraction order $n\in \mathbb{N}$. Here $\lambda$ and $a$ represent the wavelength of the incident beam and the lattice constant, respectively. For details the interested reader is referred to Ref.~\protect\cite{Goodman05}.}. We have checked that for the geometrical parameters used in the design of the experimental setup, the location of the first order diffraction peaks seen in Fig.~\ref{fig:120K_300K} are observed at positions that are consistent with the predictions obtained from the grating equation for both beam temperatures.

\smallskip
We now turn to the modeling of the diffraction patterns that were obtained experimentally, which is performed on the basis of the virtual source diffraction model outlined in Sec.~\ref{sec:diffraction_model}; see Eqs.~\eqref{eq:I_sim}--\eqref{eq:error_norm}. The solid lines that appear in Fig.~\ref{fig:120K_300K} represent the predictions of the diffraction model for the two beam temperatures considered. These results were obtained by averaging the results of \num{1000} realizations of the source. In order to produce these simulation results, the geometrical parameters characterizing the experimental setup, the sample, and the temperature (and wavelength) of the incident beam were assumed known. An overview of these parameters and their values is presented in Table~\ref{tab:simulation_parameters}.

It was found that the diffraction model produces diffraction patterns which forms are sensitive to the width of the virtual source $\sigma$, see Eq.~\eqref{eq:source_description}. The widths of the diffraction peaks strongly depend on the shape of the incoming field. By changing the width $\sigma$ in the source model, the width of the modeled diffraction peaks change. A wider Gaussian envelope used for the incident beam (a larger value for $\sigma$)  will broaden the diffraction peaks. In principle, the source width can be measured experimentally~\cite{Reisinger2007,Eder2014}. However, such measurements typically yield significant uncertainty on the source width. Therefore, we instead decided to determine this parameter by fitting the diffraction model to the experimental diffraction data based on Eq.~\eqref{eq:experimental_fit} and the cost function  Eq.~\eqref{eq:error_norm}. This means that the free parameters used in the diffraction model were the source width $\sigma$, the amplitude $\alpha$, and the background intensity $I_0$, see Eq.~\eqref{eq:experimental_fit}. In this way the diffraction model was fitted to the measured data sets with respect to the parameter set $\{\sigma,\alpha, I_0\}$. We discuss the roboustness of the fit at the end of this section.

From the results presented in Fig.~\ref{fig:120K_300K} it is observed that the diffraction model from Sec.~\ref{sec:diffraction_model} is capable of representing the measured diffraction patterns well: both when it comes to the positions of the diffraction peaks, their widths, and the relative amplitudes of the peaks. The width of the virtual source was determined in the fitting procedure to be \SI{77}{\micro\meter} [Fig.~\ref{fig:120K_300K}(a)] and \SI{52}{\micro\meter} [Fig.~\ref{fig:120K_300K}(b)] for the beam temperatures \SI{126}{\kelvin}  and  \SI{296}{\kelvin}, respectively.

It is interesting to note that the values we obtained for $\sigma$ when producing the numerical results (solid lines) shown in Fig.~\ref{fig:120K_300K} do agree well with values reported previously in the literature for cooled and room temperature He beams~\cite{Reisinger2007,Eder2014}. The measurements reported in these references were conducted using a zone plate to image directly the virtual source. The beam temperatures used in these two studies were \SI{125}{\kelvin} and \SI{320}{\kelvin}, which is close to the temperatures used in our experiments to allow for a comparison. For the measurements done at room temperature, the size of the virtual source is usually described as the sum of two Gaussian functions with different widths. The size of the virtual source for a beam of temperature \SI{320}{\kelvin} was reported in Ref.~\cite{Reisinger2007}. If the results from Ref.~\cite{Reisinger2007} is fitted using a single Gaussian in order to more closely resemble the model used in this paper, one obtains $\sigma=\SI[separate-uncertainty = true]{50+-10}{\micro\meter}$. This value should be compared to the result $\sigma=\SI{52}{\micro\meter}$ that we report for the beam temperature \SI{296}{\kelvin} obtained using the diffraction model Eqs.~\eqref{eq:I_sim}--\eqref{eq:experimental_fit} and the experimental data reported in Fig.~\ref{fig:120K_300K}(b).

For a \SI{125}{\kelvin} beam, the size of the virtual source is described with just one Gaussian. The lower level value reported in Ref.~\cite{Eder2014} corresponds to $\sigma=$\SI[separate-uncertainty = true]{100+-10}{\micro\meter}. This is larger than the value $\sigma=$\SI{77}{\micro\meter} we obtained in the modeling for the beam temperature \SI{126}{\kelvin} [Fig.~\ref{fig:120K_300K}(a)].

\smallskip
%
\begin{figure}
\centering
\includegraphics[width=\columnwidth]{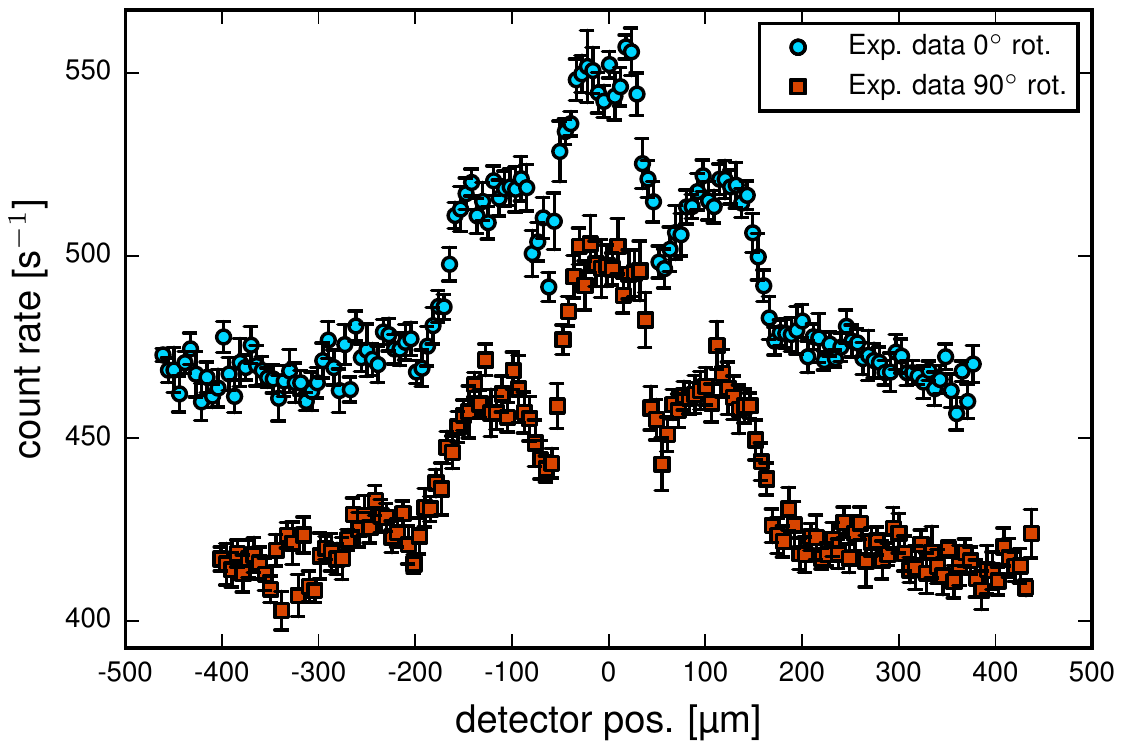}
\caption{
	\label{fig:300K_rotation}
	(Color online) Comparison of the measured transmission intensities for a helium beam temperature of \SI{296}{\kelvin}. The samples rotation angle around the $x_3$-axis was either $\phi=\ang{0}$ (circles) or $\phi=\ang{90}$ (squares). The former data set is identical to the one presented in Fig.~\ref{fig:120K_300K}. For reasons of clarity, the latter data set was shifted downwards by \num{50} counts.
}
\end{figure}
%
%
Since the grating in our sample is supposed to be square, a rotation of the sample around the $x_3$-axis by an angle of \ang{90} should, in principle, not alter the diffraction pattern that it produces. In Fig.~\ref{fig:300K_rotation} we present a comparison of two measured diffraction patterns for a beam temperature of \SI{296}{\kelvin} (room temperature); the top data set corresponds to the original position of the sample ($\phi=\ang{0}$) and the lower data set is collected after rotating the sample an angle $\phi=\ang{90}$ relative to the scanning direction. The latter data set was shifted downward by \num{50} counts per second for reasons of clarity. It is observed that the two data sets presented in Fig.~\ref{fig:300K_rotation} are consistent.

%
\begin{figure}
\centering
\includegraphics[width=\columnwidth]{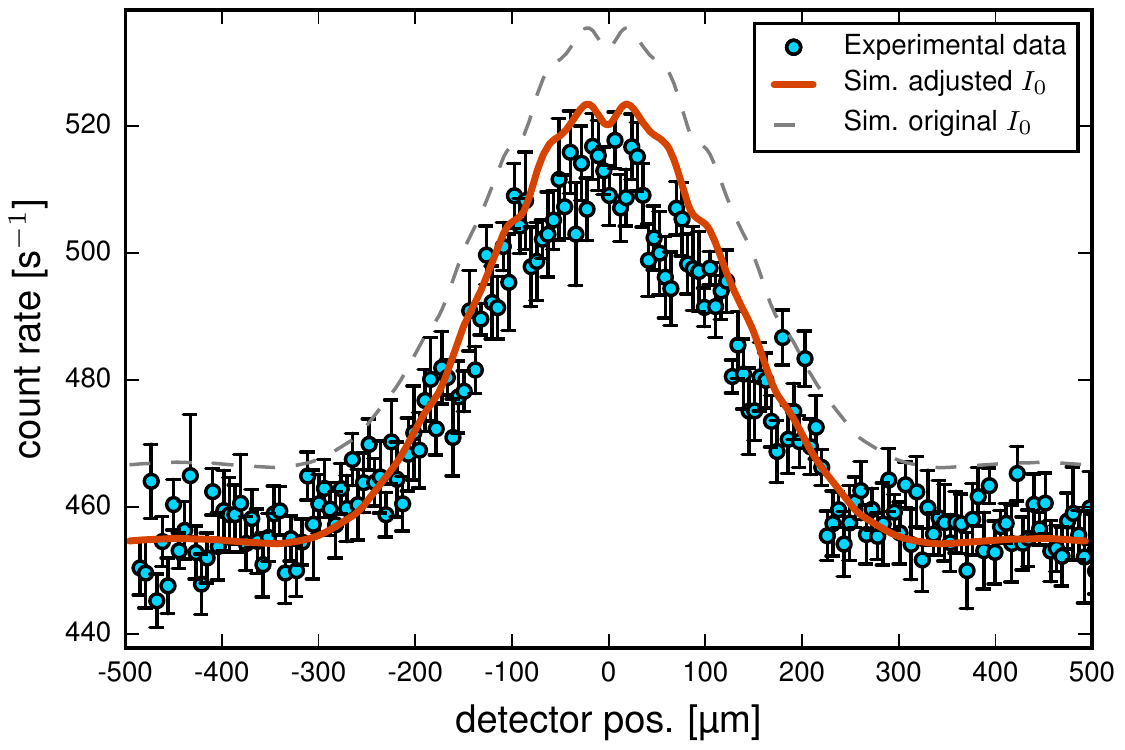}
\caption{\label{fig:300K_45}(Color online) Same as Fig.~\ref{fig:120K_300K}(b), but the photonic crystal was rotated an angle $\phi=\SI{45}{\degree}$ relative the scan direction. This is the orientation between the two curves in Fig.~\ref{fig:300K_rotation}. Parameters used for the simulation is shown in Table~\ref{tab:simulation_parameters}. The source parameters $\{\sigma,\alpha\}$ are the same as were found for the fit shown in Fig.~\ref{fig:120K_300K}(b). The dashed simulation fit also used the background $I_0$ found previously, while the solid curve uses an adjusted background.
}
\end{figure} 
%
%
The diffraction patterns are expected to change relative to what is presented in Fig.~\ref{fig:300K_rotation} if the sample is rotated an angle $\phi$ from the scan direction if this angle is not a multiple of \ang{90}. In Fig.~\ref{fig:300K_45}  the experimentally obtained mean scan-curve corresponding to a rotation angle of \ang{45} is presented. From this figure it is apparent that the obtained diffraction pattern is different from those presented in Fig.~\ref{fig:300K_rotation}, which correspond to the sample rotation of $\phi=\ang{0}$ and \ang{90}. The measured data set shown in Fig.~\ref{fig:300K_45} does not clearly display a diffraction pattern with several peaks. This is at least the case with a scan interval along the $x_1$-axis from \SI{-500}{\micro\meter} to \SI{500}{\micro\meter}. A square grid will produce a square diffraction pattern and the widths of the diffraction peaks can depend on the rotation of the sample. If the scanning direction is not properly aligned along one of the axes of the grid, the diffractive orders that were previously captured at the same position by the scanning slit might move slightly and the peaks become more diffuse. Depending on the rotation angle $\phi$, the scanning slit may pick up contributions from different diffraction peaks at different locations. 

The data measured for the rotation angle $\phi=\ang{45}$ are consistent with what is expected theoretically. To see this, we present in Fig.~\ref{fig:300K_45} the prediction of the virtual source diffraction model as a solid line, and good agreement is found between the measured and simulated data. It is important to stress that in obtaining the theoretical data, the only free parameter was a small adjustment in the background signal $I_0$. Since the source used in obtaining the measured results presented in Figs.~\ref{fig:120K_300K}(b) and \ref{fig:300K_45} is the same, the values for the source parameters $\{\sigma,\alpha\}$ should be the same. However, the value for the background $I_0$ is known to vary over time. As the measurements at different angles were performed several days apart, this value is expected to change. Therefore, the values for $\{\sigma,\alpha\}$ obtained when producing the solid line in Fig.~\ref{fig:120K_300K}(b) were assumed when producing the theoretical prediction (solid line) presented in Fig.~\ref{fig:300K_45}, with only a small adjustment to $I_0$. The dashed line in Fig.~\ref{fig:300K_45} shows the simulation when taking all parameters in the set $\{\sigma,\alpha,I_0\}$ from Fig.~\ref{fig:120K_300K}(b). The values for $\{\sigma,\alpha\}$ obtained by inversion of \textit{one} experimentally obtained scan-curve can be used to accurately predict the results for other scan directions obtained by using the same source setup. This testifies to the consistency and usefulness of the approach. 

\medskip

\begin{figure}
\centering
\includegraphics[width=\columnwidth]{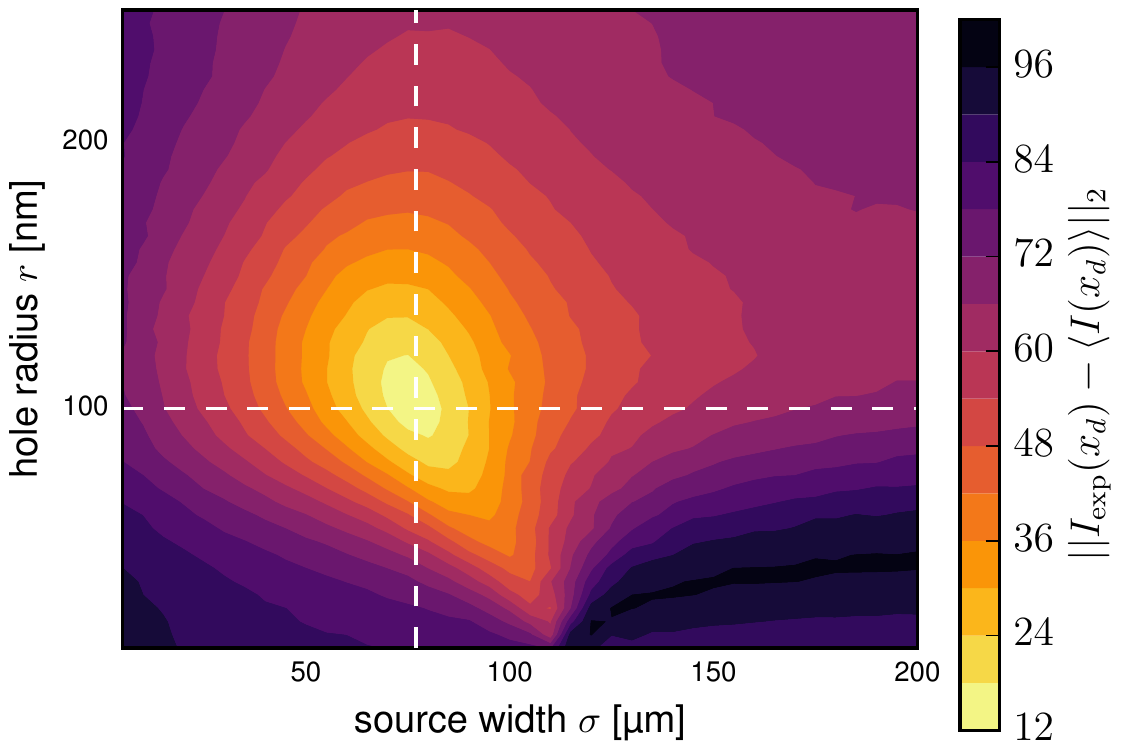}
\caption{\label{fig:rsd_scan}(Color online) Error norm shown for a large change in the source width $\sigma$ and the radius of the holes in the grid. Hole radius variation from \SI{0}{\nano\meter} to half of the period between holes, and source width $\sigma$ variation from \SI{0}{\micro\meter} to the radius of the skimmer opening. The other parameters were held constant equal to the parameters in Table~\ref{tab:simulation_parameters} for a beam temperature of \SI{126}{\kelvin}. The dashed lines indicate the parameters used for the simulation shown in Fig.~\ref{fig:120K_300K}(a) ($w$=\SI{100}{\nano\meter}, $\sigma$=\SI{77}{\micro\meter}).}
\end{figure}

\smallskip
We have explicitly demonstrated the usefulness of the virtual source diffraction model for the purpose of representing, interpreting and simulating neutral helium atom diffraction through periodic structures. Central to the approach is the determination of parameters that characterize the virtual source and, potentially, properties of the sample that are not known in advance. We now turn to the robustness and accuracy of the determination of such parameters. To this end, we present in Fig.~\ref{fig:rsd_scan} a contour plot of the cost function (or the error norm)~\eqref{eq:error_norm} for a large variation of the source width $\sigma$ and the radius of the holes. In obtaining these results the beam temperature was assumed to be \SI{126}{\kelvin}, the experimental data from Fig.~\ref{fig:120K_300K}(a) were used, while the geometrical parameters of the experimental setup were those of Table~\ref{tab:simulation_parameters}. From the results presented in Fig.~\ref{fig:rsd_scan}, a well-defined region of parameter space is observed for which the cost function is at a minimum. Moreover, this region also encompasses the known values for the radius of the holes $w$ and the previously fitted value for the source width $\sigma$. The cost function (the error) is also observed to be a smoothly varying function of $w$ and $\sigma$. This is at least the case for the region of parameter space that we considered. Such behavior of the cost function makes the determination of the unknown parameters (like $\{\sigma,w\}$) easier. 

To test the robustness of the optimization procedure, a broader search with more free parameters was also performed using the Nelder-Mead algorithm with adaptive parameters~\cite{Gao2012}. Optimizations with respect to the source width $\sigma$, the hole radius as well as the period of the grating were performed successfully using the measured scan data from Fig.~\ref{fig:120K_300K}(a).
When starting the optimization from a large random simplex covering the parameter space, we reliably found a minimum in the cost function that corresponded to parameters that were close to the values for the hole radius and the lattice constant determined from scanning electron microscopy images (see Fig.~\ref{fig:crystal}) and the source width $\sigma$ determined previously using a lower-dimensional parameter space. For instance, starting from a large and randomly chosen initial simplex, a typical fit was $\{\sigma,w,p\} = \{\SI{74}{\micro\meter},\SI{105}{\nano\meter},\SI{500}{\nano\meter}\}$. This is close to the parameters used previously.
One issue we encountered in the optimization process was that the cost function was rather flat around the minimum. Slight variations in the cost function when only taking a few realizations of the source made it hard to find exact parameters. The situation improved when increasing the number of realizations used to calculate the theoretical diffraction pattern, but at the expense of longer simulation time. This leads us to conclude that the virtual source diffraction model can be a viable tool for characterizing the average properties of photonic crystals similar to the one shown in Fig.~\ref{fig:crystal}, but that a higher number of source realizations is needed in the modeling in order to obtain reliable parameter retrival.

\section{Conclusions and Outlook}
Helium diffraction measurements from a photonic crystal structure have been presented. The diffraction patterns measured are in excellent agreement with the theoretical results obtained by including effects of a source of finite extension. The model constructed to fit the experimental data could be used as a future tool for extracting the parameters of periodic gratings on the nanometer scale. Furthermore, the model may be used to describe the behavior of helium beams, which is important for a range of applications, including the development of an efficient neutral helium microscope~\cite{Adria2016,Adria2017}.

\begin{acknowledgments}
The authors gratefully acknowledge support from the Research Council of Norway, Fripro Project 213453 and Forny Project 234159. The research of I.S. was supported in part by the Research Council of Norway Contract No. 216699 and The French National Research Agency~(ANR) under contract ANR-15-CHIN-0003-01.
\end{acknowledgments}

\bibliography{bibtex.bib}

\end{document}